\documentclass[%
 reprint,
superscriptaddress,
groupedaddress,
showpacs,preprintnumbers,
nofootinbib,
nobibnotes,
 amsmath,amssymb,
prb,
]{revtex4-1}

\usepackage{graphicx}
\usepackage{dcolumn}
\usepackage{bm}
\usepackage{amssymb}   
\usepackage{amsmath}
\usepackage{color}
\usepackage[normalem]{ulem}
\usepackage{siunitx}
\usepackage{mathrsfs}
\usepackage{tikz}
\usepackage{amsfonts}
\usepackage{kantlipsum}
\usepackage{subfigure}
\usepackage{multirow}

\newcommand{\figref}[1]{{Fig.~\ref{#1}}}
\newcommand{\tabref}[1]{{Table~\ref{#1}}}
\newcommand{\secref}[1]{{Section~\ref{#1}}}

\begin{document}


\title{Computational Analysis of Dispersive and Nonlinear 2D Materials by Using a Novel GS-FDTD Method}

\author{Jian Wei You}
\affiliation{%
 Department of Electronic and Electrical Engineering, University College London, London WC1E 7JE, United Kingdom
}%

\author{Edward Threlfall}%
\author{Dominic F. G. Gallagher}%
\affiliation{%
 Photon Design Ltd., 34 Leopold Street, Oxford OX4 1TW, United
Kingdom
}%

\author{Nicolae C. Panoiu}
\affiliation{ Department of Electronic and Electrical Engineering, University College London, London WC1E 7JE, United Kingdom.  Email: n.panoiu@ucl.ac.uk
}%

\date{\today}

\begin{abstract}
In this paper, we propose a novel numerical method for modeling nanostructures containing dispersive and nonlinear two-dimensional (2D) materials, by incorporating a nonlinear generalized source (GS) into the finite-difference time-domain (FDTD) method. Starting from the expressions of nonlinear currents characterizing nonlinear processes in 2D materials, such as second- and third-harmonic generation, we prove that the nonlinear response of such nanostructures can be rigorously determined using two linear simulations. In the first simulation, one computes the linear response of the system upon its excitation by a pulsed incoming wave, whereas in the second one the system is excited by a nonlinear generalized source, which is determined by the linear near-field calculated in the first linear simulation. This new method is particularly suitable for the analysis of dispersive and nonlinear 2D materials, such as graphene and transition-metal dichalcogenides, chiefly because, unlike the case of most alternative approaches, it does not require the thickness of the 2D material. In order to investigate the accuracy of the proposed GS-FDTD method and illustrate its versatility, the linear and nonlinear response of graphene gratings have been calculated and compared to results obtained using alternative methods. Importantly, the proposed GS-FDTD can be extended to 3D bulk nonlinearities, rendering it a powerful tool for the design and analysis of more complicated nanodevices.
\end{abstract}

\maketitle


\section{Introduction}

Since the first atomic-scale thin material (graphene) was successfully isolated
from graphite in 2004 \cite{ngm04sci}, a plethora of new two-dimensional (2D) materials have been
discovered and synthesized \cite{mlh10prl,scl10nl,vpq12prl,wkk12natnano,g15nat,blm15acsnano}.
Their novel and unique properties, combined with promising technological potential, have spurred a
tremendous research interest geared towards both fundamental science and practical applications.
For instance, graphene and transition-metal dichalcogenide (TMDC) monolayers \cite{wkk12natnano},
which are just two examples of 2D materials, have already been employed in a broad array of
applications, including electronics \cite{g09sci,s10natnano,ldj10sci,rrb11natnano}, sensors
\cite{krz14natcom}, energy storage \cite{lsc15nanoenergy}, and solar cells \cite{zxy16nanoenergy}.

In addition to their remarkable linear properties, the nonlinear optical properties of 2D
materials could play an equally important role in the development of novel active photonic devices
with new or improved functionality. For example, it has been demonstrated that third-order
nonlinear optical interactions, such as third-harmonic generation (THG)
\cite{hdp13prx,cvs14njp,csa16acsnano,yyn17ptrsa} and Kerr effect \cite{zvb12ol}, are strongly
enhanced in graphene when propagating or localized surface-plasmon polaritons (SPPs) are
generated. Moreover, second-order nonlinear optical processes, such as second-harmonic generation
(SHG) \cite{knc13prb,mab13prb,wmg15prl}, are particularly strong in graphene placed on top of a
substrate or TMDC monolayers because in these cases the 2D material system is not centrosymmetric.
These nonlinear properties of 2D materials could find exciting applications both to advanced
active photonic devices, such as nanoscale frequency mixers \cite{bsh10natphoton} and
photodetectors \cite{xml09natnano}, and to the study of more fundamental phenomena, including
spatial solitons \cite{nbn12lpr}, tunable Dirac points \cite{dym15prb}, and Anderson light
localization at the nanoscale \cite{dcm15sr}.

A key enabler of rapid developments in device applications of 2D materials is access to powerful
computational methods that can describe the physics of such 2D systems, isolated or embedded in a
3D matrix. However, since one has to describe a mixture of 2D and 3D components that share the
same physical space, one has to overcome serious challenges when traditional computational methods
are to be extended to such heterostructures. Moreover, if one considers the optical properties of
photonic structures containing 2D materials, both the linear and nonlinear induced polarizations
depend strongly on frequency \cite{b08book}, which means that the linear and nonlinear optical
response of such structures are highly dispersive. These dispersive effects can be easily modeled
in the frequency domain using several well-known numerical methods
\cite{krg97jqe,fk05jlt,wp16prb,nll16josab}, as the dispersive, anisotropic and nonlinear
polarization can be conveniently and efficiently calculated in the frequency domain. However, in
order to fully describe the optical properties of the photonic system in the frequency domain,
computations must be performed for all frequencies of interest, which can greatly increase the
computational time.

To incorporate these dispersive and nonlinear effects in time-domain methods, such as the
finite-difference time-domain (FDTD) method \cite{Taf05book}, it would generally be required to
calculate complex and computationally intensive time-domain convolution integrals
\cite{JosTaf97tap,GreTaf06OptExp, GreTaf07MWCL,Taf13book}, which would consume significant computational resources. This
drawback is particularly important as the computational time and memory requirements increase
exponentially with the physical time over which the system dynamics is determined. To overcome
this problem, several simplifications and algorithm improvements have been proposed
\cite{JosTaf97tap,GreTaf06OptExp, GreTaf07MWCL, Taf13book, vc99mtt, Sul95mtt} to model instantaneous and dispersionless nonlinear
phenomena. However, modeling optical properties of 2D materials faces additional challenges
originating from embedding a 2D structure in a 3D computational grid. Whereas nonuniform grids can
be implemented in the FDTD method, the large mismatch between the grids covering the domains
containing 2D materials and bulk components and the enormous discrepancy between the optical
wavelength and the thickness of 2D materials significantly reduces the efficiency of the FDTD
method when it is applied to such 2D-3D heterostructures.

To overcome these challenges, in this paper we extend the well-known FDTD method to the
case of optical structures containing optically nonlinear 2D materials by introducing the concept
of nonlinear generalized source (GS). Specifically, we describe the nonlinear optical response of
the 2D material \textit{via} nonlinear surface currents lying on a 2D grid, and that are specific
to the particular nonlinear optical process one wishes to study. These nonlinear currents are
determined from a first linear FDTD simulation, using the specific expression relating them to the
electric field at the fundamental frequency (FF). A second FDTD simulation, with these nonlinear
currents as excitation sources, is then performed in order to compute the nonlinear optical
response of the system. Since one only needs to know the specific functional dependence of the
nonlinear currents on the field at the FF, this new numerical method, which we call GS-FDTD, can
be applied to a broad array of nonlinear processes. The paper is organized as follows. In
\secref{sec:Method}, we describe the basic algorithm of the GS-FDTD. In addition, a general
dispersive model for the electric permittivity of the 2D material considered in this work,
\textit{i.e.} graphene, is presented. In order to illustrate the versatility and efficiency of the
proposed GS-FDTD method, we compute in \secref{sec:Results} the linear and nonlinear response of
generic graphene diffraction gratings and compare them with results obtained by using the rigorous
coupled-wave analysis (RCWA) and finite-element time domain (FETD) method. Finally, the main
results and conclusions of this study are summarized in \secref{sec:Concl}.

\section{Computational Framework of GS-FDTD}\label{sec:Method}
In this section, we present the main ideas of our computational method. Thus, we first describe
how we parameterize the frequency-dependent permittivity of 2D materials and the approach we used
to translate these dependencies to the time domain. Then, we explain how nonlinear optical
interactions are first described in the frequency domain \textit{via} nonlinear surface currents
and subsequently incorporated in the time domain formulation of our GS-FDTD method.

\subsection{Incorporating 2D Nonlinear Materials in FDTD}
We begin the description of our algorithm from the Maxwell equations. Thus, the Maxwell-Amp\`{e}re
law in the absence of free charges can be expressed as:
\begin{equation}\label{eq:MA}
\nabla  \times \textbf{\textit{H}} = {\textbf{\textit{J}}_d} + {\textbf{\textit{J}}_c}
\end{equation}
where the displacement and conduction current densities, ${\textbf{\textit{J}}_d}$ and
${\textbf{\textit{J}}_c}$, respectively, are given by:
\begin{equation}\label{eq:JdJc}
{\textbf{\textit{J}}_d} = \frac{\partial \textbf{\textit{D}}}{\partial t},
~~~~~~{\textbf{\textit{J}}_c} = \sigma \textbf{\textit{E}}.
\end{equation}
with $\sigma$ being the electric conductivity.

In the frequency domain, we can decompose the electric flux density $\textbf{\textit{D}}$ into a
linear part and a nonlinear part as follows:
\begin{equation}\label{eq:D}
{\textbf{\textit{D}}}(\omega ) = {\varepsilon _0}{\varepsilon _r}{{\textbf{\textit{E}}}}(\omega) =
{{{\textbf{\textit{D}}}}_L}(\omega) + {\textbf{\textit{P}}_{NL}}({\Omega,\omega})
\end{equation}
where ${\textbf{\textit{P}}_{NL}}({\Omega,\omega })$ is the nonlinear polarization, which depends
on the FF frequency, $\omega$, and the frequency of the higher-harmonic, $\Omega$, where
$\Omega=2\omega$ ($\Omega=3\omega$) in the case of SHG (THG), and
\begin{equation}\label{eq:Dl}
{{{\textbf{\textit{D}}}}_L}(\omega) = {\varepsilon_0}\left[ {1 + {\chi^{(1)}}(\omega)}
\right]{{\textbf{\textit{E}}}}(\omega) =
{\varepsilon_0}\varepsilon_r^{(1)}(\omega){{\textbf{\textit{E}}}}(\omega)
\end{equation}

In this equation, $\varepsilon _r^{(1)}(\omega)$ and $\chi^{(1)}(\omega)$ are the linear relative
permittivity and susceptibility of the material, respectively. Using \eqref{eq:D} and
\eqref{eq:Dl} in conjunction with \eqref{eq:JdJc}, we arrive to the expression of the current
density in the frequency domain:
\begin{equation}\label{eq:JlJnl}
{\textbf{\textit{J}}_d}(\omega ) = \textbf{\textit{J}}_d^L(\omega) +
\textbf{\textit{J}}_d^{NL}({\Omega ,\omega }) = -i\omega\left[ {{{\textbf{\textit{D}}}}_L}(\omega)
+ {\textbf{\textit{P}}_{NL}}({\Omega ,\omega })\right]
\end{equation}

The generalized current density in this equation describes both the linear and nonlinear response
of the material. Therefore, if one properly incorporates this quantity into the FDTD method, the
complete response of the optical structure can be determined. For instance, the electromagnetic contribution of 2D materials are considered in our GS-FDTD method via generalized surface currents lying on 2D Yee’s grids, thus we can avoid constructing a bulk layer to approximate a 2D material in our simulations. However, for most 2D materials this
generalized current density is frequency- and intensity-dependent. As such, if one incorporates
this generalized current density directly into the regular FDTD method \cite{Taf05book}, one needs
to compute complex time-domain convolution integrals \cite{JosTaf97tap,GreTaf06OptExp, GreTaf07MWCL,Taf13book}. This would
result in a prohibitive demand of computational time and memory resources. To overcome this
roadblock, we incorporate the linear and the nonlinear parts of the generalized current density
\eqref{eq:JlJnl} into FDTD method in two separate steps. Specifically, we first determine the
nonlinear current using a linear FDTD simulation, transform this nonlinear current in the time
domain, then, in a second linear FDTD simulation, this current is used as a generalized source of
the nonlinear field. These steps are described in detail in what follows.

\subsection{Linear Simulation}
In the linear FDTD simulation, we assume that there are only linear materials in the computational
domain, and based on this assumption we calculate the corresponding time-dependent electromagnetic
field distribution. In addition, the electric field at the location of (nonlinear) 2D materials,
which can be viewed as the field at the FF, is recorded to be used in the next step of the
algorithm, namely to evaluate the nonlinear GS currents.

The linear properties of most of 2D materials, including graphene and TMDC monolayers, are
generally frequency-dependent. To include these dispersive effects in the FDTD method, one
generally uses some well-known dispersion models, such as Debye, Drude, and Lorentz, to fit the
frequency-dependent permittivity. As a result, the dispersive medium can be simulated by employing
the auxiliary-difference-equation (ADE) FDTD method \cite{Taf05book}. However, each dispersion
model is only suitable for particular applications. For instance, the Debye model is generally
used to describe the dispersive features of human tissues and soil, the Drude model is suitable
for noble metals and plasma, and the Lorentz model is widely used to describe the optical
dispersion of semiconductors and polaritonic materials.

\begin{table}[!b]
\centering
\caption{\bf The coefficients $a_m^0$, $a_m^1$, $b_m^0$, $b_m^1$, $b_m^2$ describing
several well-known dispersion models, namely Debye, Drude, Lorentz, and modified
Lorentz.}
\begin{tabular}{c|c|c}
\hline
\multirow{2}{*}{Model} & Dispersion &  Coefficients \\
\cline{2-3}
& ${\varepsilon _m}\left( \omega \right)$ & $\frac{{a_m^0 + a_m^1\left( {-i\omega } \right)}}{{b_m^0 + b_m^1\left( {-i\omega } \right) + b_m^2{{\left( {-i\omega } \right)}^2}}}$ \\
\hline
Debye &  $\frac{{\Delta \varepsilon }}{{1 - 2i\omega {\gamma _m}}}$ &  $\begin{array}{l}
\;\;\;\;\;\;a_m^0 = \Delta \varepsilon ,\;\;\;a_m^1 = 0,\\
b_m^0 = 1,\;\;\;b_m^1 = 2{\gamma _m},\;\;b_m^2 = 0
\end{array}$ \\
\hline
Drude &  $\frac{{\Delta \varepsilon  \cdot \omega _m^2}}{{-2i\omega {\gamma _m} - {\omega ^2}}}$   & $\begin{array}{l}
\;\;a_m^0 = \Delta \varepsilon  \cdot \omega _m^2,\;\;\;\;a_m^1 = 0,\\
b_m^0 = 0,\;\;\;b_m^1 = 2{\gamma _m},\;\;\;b_m^2 = 1
\end{array}$ \\
\hline
Lorentz &  $\frac{{\Delta \varepsilon  \cdot \omega _m^2}}{{\omega _m^2 - 2i\omega {\gamma _m} - {\omega ^2}}}$
   & $\begin{array}{l}
\;a_m^0 = \Delta \varepsilon  \cdot \omega _m^2,\;a_m^1 = 0,\\
b_m^0 = \omega _m^2,\;b_m^1 = 2{\gamma _m},\;b_m^2 = 1
\end{array}$ \\
\hline Lorentz-M &  $\frac{{\Delta \varepsilon  \cdot \omega _m^2 -i\omega \Delta \varepsilon
{\gamma^{\prime}_{m}}}}{{\omega _m^2 - 2i\omega {\gamma _m} - {\omega ^2}}}$   & $\begin{array}{l}
\;a_m^0 = \Delta \varepsilon  \cdot \omega _m^2,\;a_m^1 = \Delta \varepsilon {\gamma^{\prime}_{m}},\\
b_m^0 = \omega _m^2,\;b_m^1 = 2{\gamma _m},\;b_m^2 = 1
\end{array}$ \\
\hline
\end{tabular}
  \label{tab:disp}
\end{table}

The frequency dispersion of graphene permittivity cannot be described by any of these models.
Therefore, we use a more general approach, which can be applied to practically any function
describing the frequency dispersion of the optical medium. For the sake of specificity, we present
here this approach applied to the particular case of graphene. Thus, the linear sheet conductance
of graphene (sometimes simply called conductivity) is generally given by the Kubo's formula.
Within the random-phase approximation \cite{Hans08jap,wzg15tnano}, this formula can be reduced to
the sum of inter-band and intra-band contributions. The intra-band part is given by:
\begin{equation}\label{eq:sintra}
\sigma_{intra} = \frac{{{e^2}{k_B}T\tau}}{{\pi {\hbar ^2}\left( {1-i\omega\tau} \right)}}\left[
\frac{\mu _c}{k_B T} + 2\ln \left(e^{-\frac{\mu _c}{k_B T}} + 1\right) \right]
\end{equation}
where $\mu _c$ is the chemical potential, $\tau$ is the relaxation time, $T$ is the temperature,
$e$ is the electron charge, $k_B$ is the Boltzmann constant, and $\hbar$ is the reduced Planck's
constant. Moreover, if $\mu _c \gg k_B T$, which usually holds at room temperature, the inter-band
part can be approximated as:
\begin{equation}\label{eq:sinter}
\sigma_{inter} = \frac{{i{e^2}}}{{4\pi \hbar }}\ln \left[ {\frac{{2\left| {{\mu _c}} \right| -
({\omega  + i{\tau ^{ - 1}}})\hbar }}{{2\left| {{\mu _c}} \right| + ({\omega + i{\tau^{-1}}})\hbar
}}} \right]
\end{equation}

If we assume that the effective thickness of graphene is ${h_{eff}}$, its linear relative
permittivity $\varepsilon _r^{(1)}(\omega)$ can be written as:
\begin{equation}\label{eq:lindispgraph}
\varepsilon_r^{(1)}(\omega) = 1 + \frac{i\sigma_s}{\varepsilon_0\omega h_{eff}}
\end{equation}
where $\sigma_s = \sigma_{intra}(\omega,\mu_c,\tau,T) + {\sigma _{inter}}({\omega,\mu_c,\tau,T})$.

It can be seen that the intra-band contribution to the permittivity, at THz and optical
frequencies, is similar to that of noble metals, meaning that it can be described by a Drude
model. On the other hand, the inter-band part is similar to the dispersion of a semiconductor, and
therefore it can be represented by a Lorentz model. In order to correctly account for both
contributions, we use a more general model for frequency dispersion, which is described in what
follows.

Using a small set of dispersion coefficients, the dispersion models most used in practice, namely
Debye, Drude, Lorentz, and modified Lorentz, can be described by a common formula:
\begin{equation}\label{eq:dispfit}
\varepsilon_r^{(1)}(\omega) = \varepsilon_{\infty}+\sum\limits_{m=1}^M {{\varepsilon_m}(\omega)}
\end{equation}
where $\varepsilon_{\infty}$ is the frequency-independent part of the permittivity, $M$ is the
number of dispersion terms,
\begin{equation}
{\varepsilon_m}(\omega) =
\frac{{a_m^0+a_m^1(-i\omega)}}{{b_m^0+b_m^1(-i\omega)+b_m^2{{\left(-i\omega\right)}^2}}}
\end{equation}
and $a_m^0$, $a_m^1$, $b_m^0$, $b_m^1$, and $b_m^2$ are dispersion coefficients defining the $m$th
dispersion term. The particular values of these coefficients corresponding to the main dispersion
models used in practice are given in Table \ref{tab:disp}.

Using this general dispersion model, the linear relative permittivity of graphene and other 2D
materials can be accurately fitted. Thus, we have determined the dispersion coefficients for the
particular case of graphene with ${\mu_c}=\SI{0.6}{\electronvolt}$,
$\tau=\SI{0.25}{\pico\second}$, and $T=\SI{300}{\kelvin}$, using five dispersion terms in
\eqref{eq:dispfit} (one Drude term and four Lorentz terms), the corresponding values being
presented in \tabref{tab:graph}.
\begin{table}[!b]
\centering \caption{\bf Dispersion coefficients used to fit $\varepsilon_{r}^{(1)}(\omega)$ of
graphene with ${\mu_c}=\SI{0.6}{\electronvolt}$, $\tau=\SI{0.25}{\pico\second}$, and
$T=\SI{300}{\kelvin}$.}\label{tab:graph}
\begin{tabular}{c|c|c|c|c|c}
\hline
\multirow{2}*{$\varepsilon_\infty=1$} &$a_m^0$  &$a_m^1$ &$b_m^0$ &$b_m^1$ &$b_m^2$ \\
&($10^{30}$)    &($10^{15}$[s])  &($10^{30}$)  &($10^{15}$[s])    &(s$^2$)\\
\hline
$m$=1 &22.8    &0  &0  &3.91    &1\\
\hline
$m$=2 &1.23    &11.5    &0.37    &4.48    &1\\
\hline
$m$=3 &0.47    &3.16$\times 10^{-5}$    &0.13    &9.39    &1 \\
\hline
$m$=4 &7.56    &6.44    &3.78    &0.74    &1 \\
\hline
$m$=5 &2.05    &5.59    &1.02    &2.4    &1\\
\hline
\end{tabular}
\end{table}

\begin{figure}[t!]
\centering
\includegraphics[scale=1]{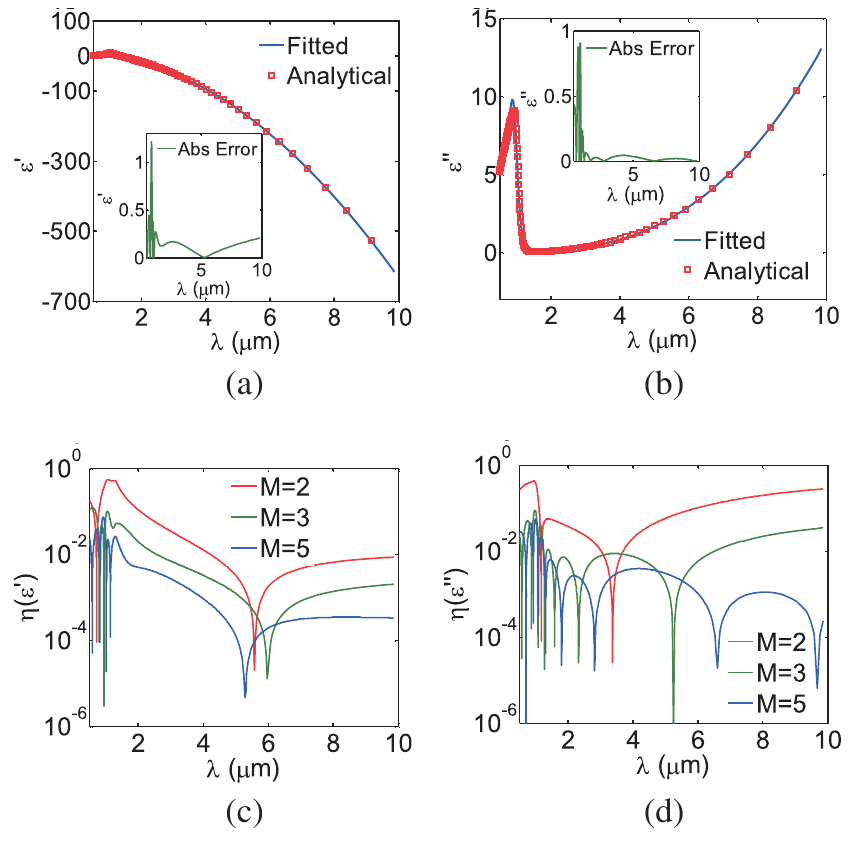}
\caption{Relative permittivity of graphene \eqref{eq:lindispgraph} fitted with the general
dispersion model \eqref{eq:dispfit}, whose coefficients are listed in \tabref{tab:graph}. (a), (b)
Error analysis of real and imaginary parts of $\varepsilon_{r}^{(1)}(\omega)$, respectively. The
insets show the absolute error. (c), (d) Convergence analysis of the relative error, where $M$ is
the number of dispersion terms.}\label{fig:disp}
\end{figure}

The data presented in Figs.~\ref{fig:disp}a and \ref{fig:disp}b show that there is a good
agreement between the analytical formula and fitting results. Moreover, one can see that the
linear permittivity of graphene at wavelengths larger than \SI{2}{\micro\meter} steeply decreases
(real part) or increases (imaginary part), which is a typical feature of permittivity of metals.
On the other hand, graphene permittivity for $\lambda<\SI{2}{\micro\meter}$ is no longer
monotonously dependent on wavelength, a common feature of semiconductors and polaritonic
materials. Additionally, the results plotted in Figs.~\ref{fig:disp}c and \ref{fig:disp}d suggest
that the maximum relative error is within $10~\%$, if five dispersion terms are used. Here, the
relative error is defined as $\vert \varepsilon_{r}^{(1)}(\omega)-\varepsilon_{fit}(\omega) \vert
/\vert \varepsilon_{r}^{(1)}(\omega) \vert$, where $\varepsilon_{fit}(\omega)$ are the fitted
values. Note that in order to achieve good fitting a relatively large number of dispersion terms
must be included, which means that simply fitting the graphene dispersion with a Drude or
Drude-Lorentz function can lead to large computational errors.

Based on \eqref{eq:Dl}, \eqref{eq:JlJnl}, and \eqref{eq:dispfit}, the frequency-dependent form of
the linear current density, $\textbf{\textit{J}}_d^L(\omega)$ can be evaluated as:
\begin{align}
\textbf{\textit{J}}_d^L(\omega) =& -i\omega {{{\textbf{\textit{D}}}}_L}(\omega)
= -i\omega{\varepsilon_0}\varepsilon_r^{(1)}(\omega){{\textbf{\textit{E}}}}(\omega) \nonumber \\
=& -i\omega\varepsilon_0 \left[\varepsilon_\infty + \sum\limits_{m = 1}^M
\varepsilon_m(\omega)\right] \textbf{\textit{E}}(\omega) = \sum\limits_{m=0}^{M}
\textbf{\textit{J}}_{m}(\omega)
\end{align}
where
$\textbf{\textit{J}}_{0}(\omega)=-i\omega\varepsilon_0\varepsilon_\infty\textbf{\textit{E}}(\omega)$
and
\begin{equation}\label{eq:Jm}
{\textbf{\textit{J}}_m}(\omega) = {\varepsilon_0}\frac{{a_m^0(-i\omega) +
a_m^1{{(-i\omega)}^2}}}{{b_m^0 + b_m^1(-i\omega) +
b_m^2{{(-i\omega)}^2}}}\textbf{\textit{E}}(\omega), ~~~m\geq1.
\end{equation}

By using the ADE method \cite{Taf05book}, the frequency-domain equation \eqref{eq:Jm} can be cast
into the following time-domain iterative relation:
\begin{equation}\label{eq:JmTD}
\textbf{\textit{J}}_m^{n+1} = c_m^0\textbf{\textit{J}}_m^n + c_m^1\textbf{\textit{J}}_m^{n-1} +
c_m^2{\textbf{\textit{E}}^{n+1}} + c_m^3{\textbf{\textit{E}}^n} +
c_m^4{\textbf{\textit{E}}^{n-1}}
\end{equation}
where the superscript $n$ indicates the $n$th time-step and the coefficients $c_{m}$'s for the
$m$th dispersion term are given by:
\begin{subequations}\label{eq:cs}
\begin{align}
c_m^0 &= \delta\left[ 2(\Delta t)^2 b_m^0 - 4b_m^2 \right], \\
c_m^1 &= \delta(b_m^1\Delta t + 2b_m^2), \\
c_m^2 &= -\delta\varepsilon_0 (a_m^0\Delta t - 2a_m^1), \\
c_m^3 &=  -4\delta\varepsilon_0a_m^1, \\
c_m^4 &= \delta\varepsilon_0(a_m^0\Delta t + 2a_m^1).
\end{align}
\end{subequations}
where $\delta=1/(b_m^1\Delta t - 2b_m^2)$ and $\Delta t$ is the time-step used in the FDTD method.
If one substitutes \eqref{eq:JmTD} and \eqref{eq:cs} into \eqref{eq:MA}, one obtains the FDTD
iteration for the linear simulation of dispersive 2D materials as:

\begin{align}\label{eq:linFDTD}
\textbf{\textit{E}}^{n+1} =& {\beta_1}{\textbf{\textit{E}}^n} +{\beta_2}\left[\nabla\times\textbf{\textit{H}}^{n+\frac{1}{2}}\right. \nonumber \\
-& \left.\sum\limits_{m=1}^M\left(\beta_3^m\textbf{\textit{J}}_m^n+\beta_4^m\textbf{\textit{J}}_m^{n-1}\right)
-{\beta_5}\textbf{\textit{E}}^{n-1}\right]
\end{align}
where
\begin{align}
\beta_1&=\frac{\alpha_1-\alpha_2\sum\limits_{m=1}^{M}c_m^3}{1+\alpha_2\sum\limits_{m=1}^{M}c_m^2},\;\;\;
\beta_2=\frac{\alpha_2\sum\limits_{m=1}^{M}c_m^3}{1+\alpha_2\sum\limits_{m=1}^{M}c_m^2}, \nonumber \\
\beta_3^m &= \frac{c_m^0 + 1}{2},\;\;\beta_4^m=\frac{c_m^1}{2},\;\;\beta_5^{m}=\frac{1}{2}\sum\limits_{m=1}^{M}c_m^4, \nonumber \\
\alpha_1 &= \frac{2\varepsilon_0\varepsilon_\infty - \sigma\Delta
t}{2\varepsilon_0\varepsilon_\infty + \sigma\Delta t},\;\;\;{\alpha_2} = \frac{\Delta
t}{2\varepsilon_0\varepsilon_\infty + \sigma\Delta t}. \nonumber
\end{align}
In these definitions, $\sigma$ is the bulk conductivity of bulk components of the photonic
structure.

The basic steps for the linear simulation of 2D materials can be briefly summarized as follows:
\textbf{\textit{Step~1}}, update \eqref{eq:linFDTD} to compute the field ${E^{n+1}}$ at the new
time-step; \textbf{\textit{Step~2}}, calculate $J_m^{n+1}$ in \eqref{eq:JmTD} by using ${E^{n+1}}$
obtained at \textit{Step~1}; \textbf{\textit{Step~3}}, let $n=n+1$ then repeat \textit{Step~1} and
\textit{Step~2} until the energy in the entire computational region converges \cite{ytc14mtt}.

\subsection{Nonlinear Simulation}\label{sec:nlsim}
Similar to the case of bulk optical media, the nonlinear optical properties of 2D materials are
generally determined by the symmetry properties of their atomic lattice and quantified
\textit{via} nonlinear susceptibility tensors. In particular, graphene lattice belongs to the
$\mathcal{D}_{6h}$ point symmetry group, so that SHG is forbidden in a uniform graphene sheet.
However, if graphene is placed on top of a substrate, the centrosymmetric property is not
preserved because the up-down mirror symmetry is broken at the interface containing graphene, the
point symmetry group in this case being $\mathcal{C}_{6v}$. As a result, considerable SHG can be
observed in this case \cite{dh09apl,m11prb,ard14prb,dh10prb,chc16NatPhy}. Moreover, strong THG in
graphene can also occur \cite{hdp13prx,cvs14njp,csa16acsnano}, as its third-order susceptibility
is particularly large. Importantly, our method can be applied to other 2D materials, too, as it
only requires the knowledge of the nonlinear optical conductivity describing the particular
nonlinear process.

The nonlinear properties of graphene are quantified by a nonlinear surface conductivity tensor,
$\bm{\sigma}_s^{(n)}(\Omega;\omega)$, where $n$ indicates the order of the nonlinear optical
interaction. In the case of SHG, the second-order surface conductivity tensor only has three
independent nonzero components, $\sigma_{s,\perp\perp\perp}^{(2)}$,
$\sigma_{s,\parallel\parallel\perp}^{(2)}=\sigma_{s,\parallel\perp\parallel}^{(2)}$, and
$\sigma_{s,\perp\parallel\parallel}^{(2)}$, where the symbols ``$\perp$'' and ``$\parallel$''
refer to the directions perpendicular onto and parallel to the plane of graphene, respectively.
The values of these parameters used in this paper are:
$\sigma_{s,\perp\perp\perp}^{(2)}=-i\SI{9.71e-16}{\ampere\meter\per\square\volt}$,
$\sigma_{s,\parallel\parallel\perp}^{(2)}=\sigma_{s,\parallel\perp\parallel}^{(2)}=-i\SI{2.65e-16}{\ampere\meter\per\square\volt}$,
and $\sigma_{s,\perp\parallel\parallel}^{(2)}=-i\SI{2.09e-16}{\ampere\meter\per\square\volt}.$
\cite{ard14prb,dh10prb}

In the case of THG, the third-order nonlinear conductivity tensor, $\bm{\sigma}_s^{(3)}(\Omega;\omega)$, is described by a single scalar function
$\sigma_s^{(3)}(\Omega;\omega)$, \textit{via} the relation
$\bm{\sigma}_{s,ijkl}^{(3)}=\sigma_{s}^{(3)} \Delta _{ijkl}$. The function $\Delta _{ijkl}=
(\delta _{ij} \delta _{kl}+ \delta _{ik}\delta _{jl}+ \delta _{il}\delta _{jk})/3$, where
$\delta_{ij}$ is the Kronecker delta, whereas the scalar function $\sigma_s^{(3)}(\Omega;\omega)$
is given by the following expression \cite{hdp13prx,cvs14njp,csa16acsnano}:
\begin{equation}\label{eq:snonl}
\sigma_s^{(3)}(3\omega;\omega) = \frac{i\sigma_0{\left(\hbar v_F e\right)}^2}{48\pi
{\left(\hbar\omega\right)}^4}T\left(\frac{\hbar\omega}{2\left|\mu_c\right|} \right)
\end{equation}
where $v_F \approx c/300$ is the Fermi velocity, $\sigma_0 = e^2/(4\hbar)$ is the universal
dynamic conductivity of graphene, $T(x) = 17G(x) - 64G(2x) + 45G(3x)$, and
$G(x)=\ln\vert(1+x)/(1-x)\vert+i\pi H(|x|-1)$, $H(x)$ being the Heaviside step function.

The nonlinear surface conductivity and nonlinear bulk susceptibility,
$\bm{\chi}^{(n)}(\Omega;\omega)$, define the nonlinear current, $\textbf{\textit{J}}_d^{NL}$, and
nonlinear polarization, $\textbf{\textit{P}}_{NL}$, respectively. Thus, in the SHG case, these
physical quantities are determined by the relations:
\begin{subequations}\label{eq:JPnlSHG}
\begin{align}
\textit{J}_{d,i}^{NL}(\Omega,\omega) &=
\sum\limits_{jk}\sigma_{s,ijk}^{(2)}(\Omega;\omega)\textit{E}_j(\omega)\textit{E}_k(\omega),\label{eq:JnlSHG}\\
\textit{P}_{i}^{NL}(\Omega,\omega) &= \varepsilon_0 \sum\limits_{jk}
\chi_{ijk}^{(2)}(\Omega;\omega)\textit{E}_j(\omega)\textit{E}_k(\omega). \label{eq:PnlSHG}
\end{align}
\end{subequations}
whereas in the THG case they are given by:
\begin{subequations}\label{eq:JPnlTHG}
\begin{align}
\textit{J}_{d,i}^{NL}(\Omega,\omega) &=
\sum\limits_{jkl}\sigma_{s,ijkl}^{(3)}(\Omega;\omega)\textit{E}_j(\omega)\textit{E}_k(\omega)\textit{E}_l(\omega),\label{eq:JnlTHG}\\
\textit{P}_{i}^{NL}(\Omega,\omega) &= \varepsilon_0 \sum\limits_{jkl}
\chi_{ijkl}^{(3)}(\Omega;\omega)\textit{E}_j(\omega)\textit{E}_k(\omega)\textit{E}_l(\omega).\label{eq:PnlTHG}
\end{align}
\end{subequations}
where, the subscript indices $i,j,k,l=x,y,z$. Using the relation
$\textbf{\textit{J}}_d^{NL}(\Omega,\omega)=-i\omega\textbf{\textit{P}}_{NL}(\Omega,\omega)$ in
conjunction with \eqref{eq:JPnlSHG} and \eqref{eq:JPnlTHG}, and keeping in mind that
$\textbf{\textit{J}}_d^{NL}(\Omega,\omega)$ is a surface current, one can easily prove that
$\bm{\chi}^{(n)}(\Omega;\omega)=[i/(\varepsilon_0\Omega
h_{eff})]\bm{\sigma}_s^{(n)}(\Omega;\omega)$.

By contrasting \eqref{eq:Jm} with \eqref{eq:JnlSHG} and \eqref{eq:JnlTHG}, it can be seen that it is
fairly simple to cast the linear current \eqref{eq:Jm} into a time-domain iteration relation by
using the ADE method, due to its linear field dependence feature and the rational polynomial
format of the dispersion model. By contrast, the time-domain expressions of dispersive and
intensity-dependent nonlinear currents \eqref{eq:JnlSHG} and \eqref{eq:JnlTHG} require the
calculation of complex, multiple time-domain convolution integrals. Specifically, the time-domain
convolution integral corresponding to \eqref{eq:JnlTHG} is written as:
\begin{align}\label{eq:convnonl}
J_{d,i}^{(3)}(t) =&
\sum_{jkl}\int_{-\infty}^{\infty}\int_{-\infty}^{\infty}\int_{-\infty}^{\infty}
\left[\sigma_{s,ijkl}^{(3)}(t-\tau_1,t-\tau_2,t-\tau_3)\right. \nonumber \\
&\left. \times E_j(\tau_1)E_k(\tau_2)E_l(\tau_3)\right]d\tau_{1}d\tau_{2}d\tau_{3}
\end{align}

In addition, this convolution integral describes not only THG processes but a multitude of other
nonlinear optical interactions that might not be of interest for the particular problem under
investigation.

To understand how these problems can be circumvented, let us first remind the reader
that, owing to the leap-frog nature of the FDTD iterative calculations, in order to march in time
the corresponding iterative relations one only needs to store the fields at the current time-step,
$n\Delta t$, and the next time-step, $(n+1)\Delta t$. This means that only $2 \times 3 \times M$
electric field values are required to be stored, which correspond to 2 different time-steps, 3
field components, and $M$ grid points. Even in the dispersive case \eqref{eq:linFDTD}, one only
needs to save $3 \times 3 \times M$ electric field values, that is 3 different time-steps, namely
the previous time-step, $(n-1)\Delta t$, current, and next time-step. On the other hand, due to
the non-instantaneous response of the medium implied by \eqref{eq:convnonl}, the electric field at
all past time-steps must be stored in order to be able to calculate the nonlinear current density
at the next time-step, $(n+1)\Delta t$. In other words, we need to store $n \times 3 \times M$
electric field values at the time-step $n\Delta t$. This is a challenge in traditional FDTD method,
as the memory resources and computational time required to compute \eqref{eq:convnonl} would
rapidly increase with the number of time steps. In order to overcome this challenge, several
solutions have been proposed \cite{GreTaf06OptExp,GreTaf07MWCL,Taf13book,vc99mtt,Sul95mtt}, most of them aiming to
simplify the calculation of \eqref{eq:convnonl} by employing certain assumptions. Different from
these previous works, in our approach we augment the standard FDTD framework with a generalized
source method, eliminating in this process the need to calculate the time-domain convolution
integral \eqref{eq:convnonl}. This novel GS-FDTD method is detailed in the next subsection.

\subsection{GS-FDTD Method}
Second- and third-harmonic generation are nonlinear optical processes pertaining to three- and
four-wave interactions in nonlinear optical media, respectively. They occur when two (SHG) or
three (THG) photons with the same frequency $\omega_0$ combine and generate a photon with
frequency $2\omega_0$ (SHG) or $3\omega_0$ (THG), respectively. These nonlinear optical processes
are determined by the local field at the fundamental frequency $\omega_0$. Importantly, other
nonlinear processes are possible, such as sum- and difference-frequency generation or four-wave
mixing, and one key feature of our numerical method is that it allows one to isolate the nonlinear
optical interaction of interest and disregard all the others. This is a particularly important
feature because the method is formulated in the time-domain, which generally makes it difficult to
study only a specific nonlinear optical process. Our method is ideally suited for such studies
because we can selectively separate a certain nonlinear optical interaction by implementing the
nonlinear simulation as two separate linear FDTD simulations. In the first linear simulation the
excitation is a regular linear source, such as a plane-wave excitation, whereas in the second
linear simulation the excitation is a nonlinear generalized source. This nonlinear generalized
source is fully determined by the specific nonlinear optical process that is investigated, and
thus one can readily separate specific nonlinear interactions from the multitude of possible
nonlinear effects. The implementation of the proposed method is divided in the following three
steps.

\textbf{\textit{Step 1: Linear simulation at FF.}} In the first linear FDTD
simulation, we assume that there are only linear materials in the computational region, and excite
this linear system at the FF with a linear source, such as a plane-wave or a voltage source. As
previously explained, we can calculate the time-domain near-field distribution within a frequency
range of interest by using a single FDTD simulation.

\textbf{\textit{Step 2: Nonlinear GS evaluation.}} Before performing the second
linear FDTD simulation, we evaluate the GS that will be used in the second linear FDTD simulation
using \eqref{eq:JnlSHG} and \eqref{eq:JnlTHG}. Specifically, the nonlinear current density is
determined first in the frequency domain using the near-field calculated at a series of
fundamental frequencies. More specifically, the time-domain near-field distribution at the FF
obtained at \textit{Step 1} is transformed into the frequency domain using the discrete-time
Fourier transformation (DTFT). Subsequently, we substitute these frequency domain near-fields into
\eqref{eq:JnlSHG} and \eqref{eq:JnlTHG} to evaluate the nonlinear current density. In order to
incorporate these nonlinear current sources into the FDTD simulation, an inverse DTFT is applied
to transform these frequency-domain nonlinear current sources into the time domain. It should be
noted that the number of frequency sampling points in above DTFTs should strictly satisfy the
Nyquist-Shannon sampling theorem, so that the time-domain nonlinear current source can be
recovered accurately \textit{via} the inverse DTFT. This nonlinear current source only depends on
the electric field at FF.

\textbf{\textit{Step 3: Linear simulation at high-order frequency.}} In the second linear FDTD
simulation performed after the first one has completed, we again assume that the whole
computational region contains only linear optical materials. However, unlike the first linear FDTD
simulation performed at \textit{Step 1}, in the second linear FDTD simulation the excitation
source is the time-dependent nonlinear current source obtained at \textit{Step 2}. In this way, we
can accurately model the nonlinear interactions between arbitrary incident electromagnetic waves
and photonic structures containing nonlinear 2D materials.

It should be noted that as sources in the second linear FDTD simulation one can simultaneously use
both the linear and nonlinear sources, in order to ensure that the computational setup more
closely replicates real-world experiments. However, in our previous work
\cite{wp16prb,bp10prb,yzgcc18TEMC,ywz15ted,ywc14ted,yyn17ptrsa,ywc14mwcl}, we found out that the nonlinear
response is extremely weak as compared to the strong linear excitation signal. As a result, once
we introduce the linear source into the second FDTD simulation the nonlinear signal becomes buried
into the noise spectrum of linear excitation signal. For this reason, as excitation in the second
linear FDTD simulation we only use the nonlinear current source. Equally important, the fact that
the nonlinear signal is much weaker than the linear one ensures that the down-conversion process
from higher-harmonics to the FF can be neglected (also known as the undepleted pump
approximation), which means that the only approximation contained in our approach is valid.

Compared to frequency-domain methods, the electric field at different frequencies in
\eqref{eq:JnlSHG} and \eqref{eq:JnlTHG} can be obtained from a single FDTD simulation \textit{via}
DTFT, rather than repeating the simulation for each frequency. Thus, it is expected that the
GS-FDTD is generally faster than nonlinear, frequency-domain methods, particularly when the
nonlinear response of the system is required within a broad spectral range.

\section{Results and Discussion}\label{sec:Results}
The proposed GS-FDTD method is a general numerical approach to study nonlinear optical effects,
such as SHG and THG, in 2D materials. In order to illustrate its versatility and efficiency, we
investigate here a double resonance phenomenon \cite{yyn17ptrsa} in photonic nanostructures made
of graphene, which is a typical dispersive and nonlinear 2D material. In the following
simulations, the frequency-domain FEM results are calculated by CST Microwave Software
\cite{cstsoftware}, and the time-domain FEM (FETD) results are obtained by using OmniSim/FETD simulator
\cite{pdsoftware}. The FDTD, GS-FDTD, RCWA, and GS-RCWA results are calculated using our in-house
developed codes.

\subsection{Geometry of the Optical Structure}
\begin{figure}[t!]
\centering
\includegraphics[scale=0.34]{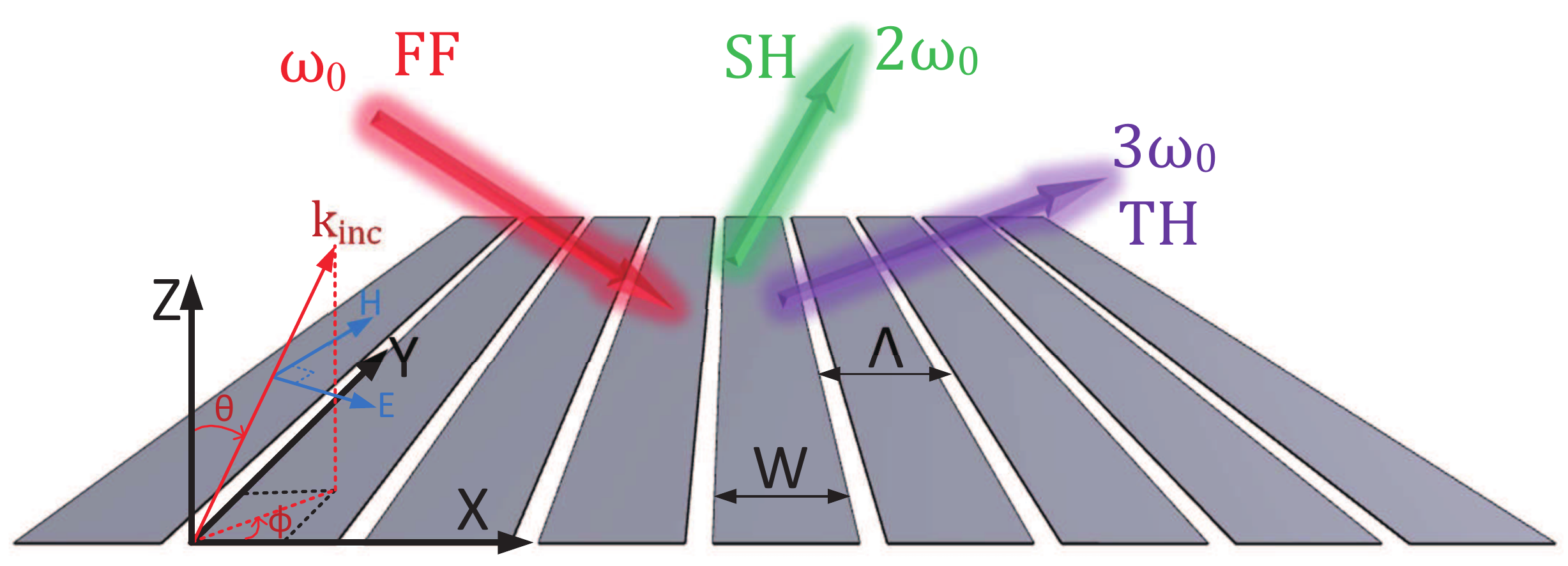}
\caption{Schematic of a graphene grating with period, $\Lambda$, and width of graphene ribbons,
$W$.}\label{fig:geom}
\end{figure}

\begin{figure}[t!]
\centering
  \includegraphics[scale=0.34]{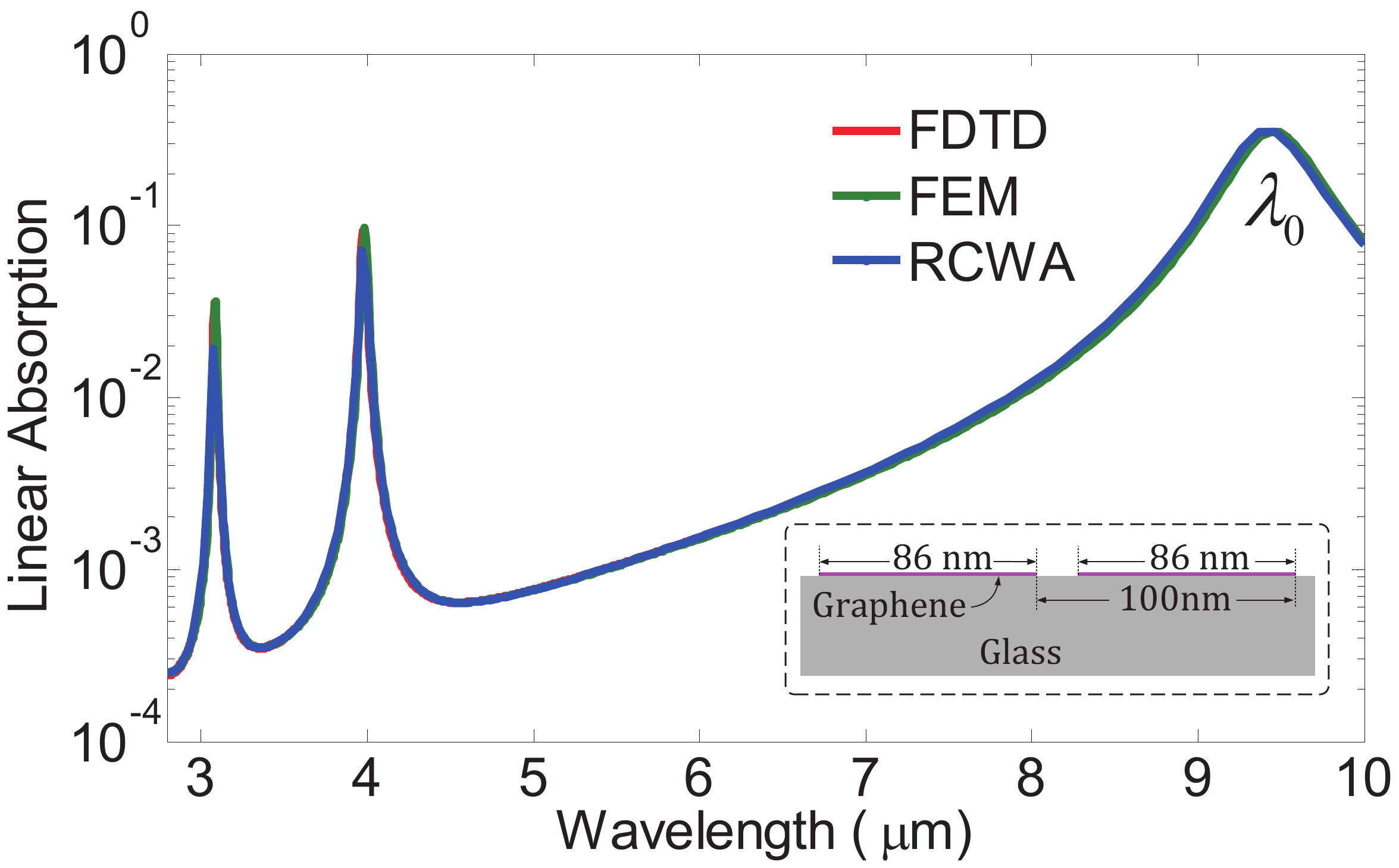}
\caption{Comparison of the absorption spectra of a graphene grating on a glass substrate,
calculated by different methods.}\label{fig:GlassSubFF}
\end{figure}

As schematically shown in \figref{fig:geom}, the studied structure is a graphene optical grating
consisting of a periodic distribution of graphene ribbons oriented along the $x$-axis. In this
example, the period is $\Lambda=\SI{100}{\nano\meter}$ and the width of the graphene ribbons is
$W=\SI{86}{\nano\meter}$. The graphene grating lies in the $xy$-plane, and in the THG case it is
assumed to be in a suspended membrane configuration. On the other hand, in the SHG case the
graphene grating is deposited on a glass substrate with $\varepsilon_r=2.25$, as per the inset of
\figref{fig:GlassSubFF}. The linear properties of graphene are described by its linear surface
conductivity as given by \eqref{eq:sintra} and \eqref{eq:sinter}. In the following simulations,
the chemical potential of graphene is $\mu_c=\SI{0.6}{\electronvolt}$, the relaxation time
$\tau=\SI{0,25}{\pico\second}$, and the temperature $T=\SI{300}{\kelvin}$. Moreover, the
third-order nonlinear optical response of graphene is characterized by its third-order surface
conductivity as expressed in \eqref{eq:snonl}, whereas the three independent components of the
second-order susceptibility tensor are provided in Section 2\ref{sec:nlsim}.

In the following examples, the graphene grating is illuminated at fundamental frequencies by a
plane wave. This plane wave carries a Gaussian pulse, which covers the fundamental-frequency
domain ranging from \SIrange{30}{150}{\tera\hertz}. The angles defining the incidence direction
are $\theta=\pi$ and $\phi=0$ (see \figref{fig:geom}), namely the grating is illuminated by a
normally incident plane wave polarized along the $x$-axis.

\subsection{Linear Results and Discussion}
\begin{figure}[t!]
\centering
\includegraphics[scale=1]{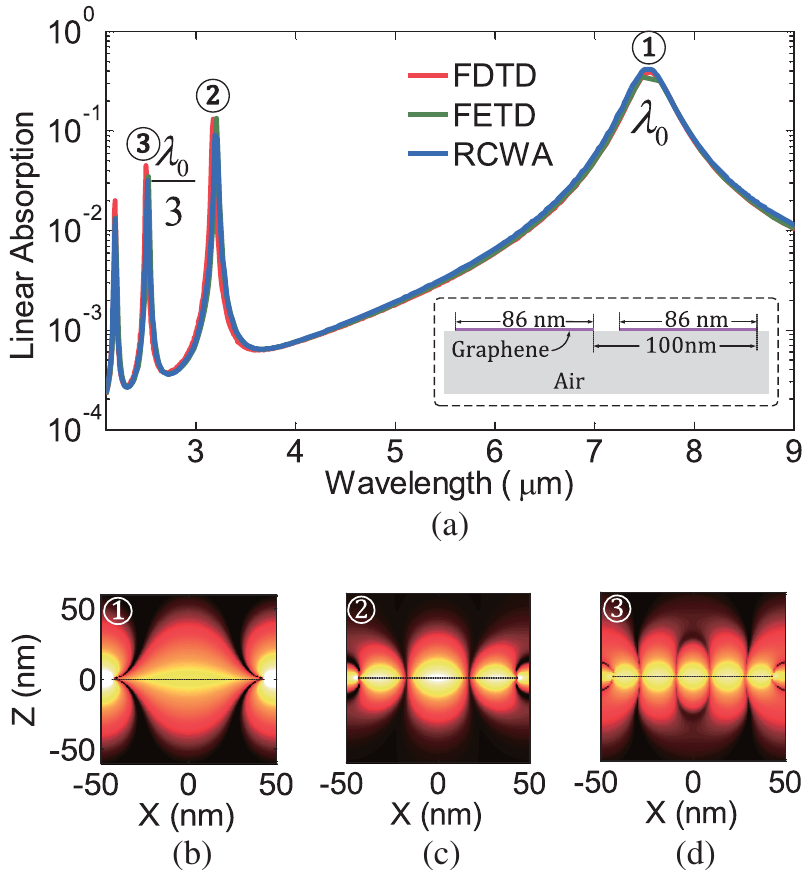}
\caption{(a) Comparison of the absorption spectra of a suspended graphene grating, calculated by
different methods. (b), (c), (d) Spatial distribution of $|E_{x}|$ corresponding to the first
three plasmon resonance modes.}\label{fig:linsp}
\end{figure}

To generate the nonlinear current sources at SH and TH, we first launch in each case a
linear simulation to obtain the near-field distribution at FF, for all frequencies of interest. To
this end, we calculated the linear optical response of the two graphene gratings using the
modified FDTD method described in \secref{sec:Method}, the corresponding results being depicted in
\figref{fig:GlassSubFF} (SHG) and \figref{fig:linsp} (THG).

These simulations reveal several important results. First, in both cases the absorption spectra
possess a series of resonances whose nature can be understood from the profile of the near-field.
These field profiles, determined for the first three resonances of the suspended graphene grating,
are plotted in Figs.~\ref{fig:linsp}b--\ref{fig:linsp}d. The strong field confinement of the
optical near-field observed at these resonance wavelengths suggests that they are the result of
excitation of localized surface plasmons on the graphene ribbons. At these resonances the local
field is strongly enhanced, which results in increased optical absorption. This behavior is
observed in both gratings, the only difference being that the presence of the dielectric substrate
induces a red-shift of the resonance wavelength.

A second phenomenon illustrated by \figref{fig:linsp} is the existence of a TH double resonance
\cite{yyn17ptrsa}. To be more specific, for the particular values of the grating parameters chosen
in this example, there are plasmon resonances both at the fundamental wavelength $\lambda_0$ and
at the TH wavelength, $\lambda_0/3$. Consequently, the near-field at both the FF and TH is
resonantly enhanced, such that one expects that the nonlinear currents at the TH are strongly
enhanced, too, as per \eqref{eq:JnlTHG}. These nonlinear currents can in turn efficiently radiate
into the continuum, which makes these specially engineered optical grating particularly effective
nonlinear optical devices for THG \cite{yyn17ptrsa}.

To verify the accuracy of our modified FDTD method, these two examples have also been
simulated by two different numerical methods, namely by OmniSim/FETD (finite-element time-domain)
\cite{pdsoftware} and RCWA \cite{wp16prb}, both using true 2D models of the graphene. The
comparison of the absorption spectra calculated using these three methods shows that there is a
very good agreement among the corresponding results, as seen both in \figref{fig:GlassSubFF} and
\figref{fig:linsp}a. This proves that our modified FDTD method is effective and accurate.

\subsection{Nonlinear Results and Discussion}
\begin{figure}[t!]
\centering
\includegraphics[scale=1]{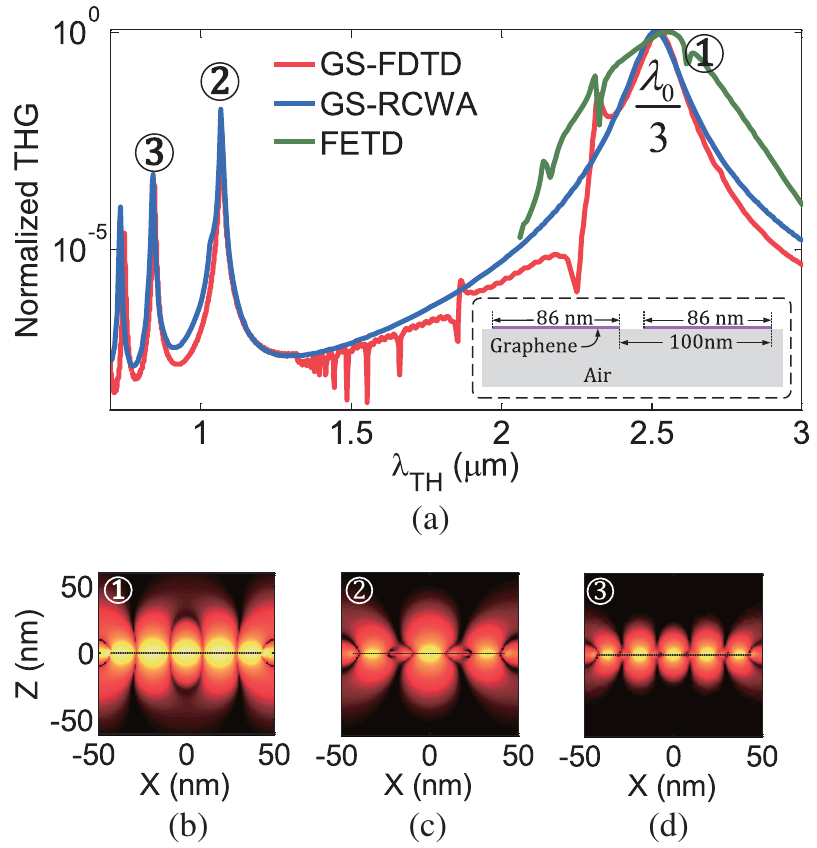}
\caption{(a) Comparison of THG spectra of suspended graphene grating calculated by three
different methods. (b), (c), (d) Spatial distribution of $|E_{x}|$ at TH corresponding to the
first three plasmon resonance modes.}\label{fig:nonlinsp}
\end{figure}

We now consider the nonlinear optical response of the two graphene gratings. Thus, the THG
spectrum of the suspended grating is shown in \figref{fig:nonlinsp}a, together with the spectra
obtained using two alternative methods. For completeness, we also present in
Figs.~\ref{fig:nonlinsp}b--\ref{fig:nonlinsp}d the near-field distributions corresponding to the
first three peaks in the THG spectrum. Similar to the linear case, the nonlinear spectrum
possesses a series of resonances, which can be mapped one-to-one to the resonances of the linear
spectra. More exactly, the peaks in THG spectrum occur at exactly a third of the resonance
wavelengths of the corresponding absorption peaks. The reason for this is that the absorption and
THG intensity are both directly determined by the local near-field at the FF. On the other
hand, the field profiles of plasmon resonances are mainly determined by the intrinsic
electromagnetic properties of graphene and the structure of the diffraction grating. Consequently,
the field profiles at specific resonance wavelengths are generally different from their linear
counterparts, which can be readily seen by comparing Fig.~\ref{fig:linsp}b and
Fig.~\ref{fig:nonlinsp}b.

As in the linear case, we also compared our results with the predictions of two alternative
methods, a GS-RCWA method introduced in \cite{wp16prb} and OmniSim/FETD \cite{pdsoftware}, which
both incorporate third-order nonlinearities. It should be noted that the latter method does not
incorporate the frequency dispersion of the nonlinear susceptibility and models graphene as a slab
with thickness of \SI{1.1}{\nano\meter} but, on the other hand, it does not rely on the undepleted
pump approximation. The results of these simulations suggest that there is a rather good agreement
among the predictions of these methods, except for some extra spectral features that are missing
in the spectrum calculated using GS-RCWA. A careful inspection of the location of these spectral
dips shows that they are due to the excitation of surface plasmons in the grating, which suggests
that the GS-RCWA method underestimates the optical loss in graphene.
\begin{figure}[!t]
  \centering
  \includegraphics[scale=0.4]{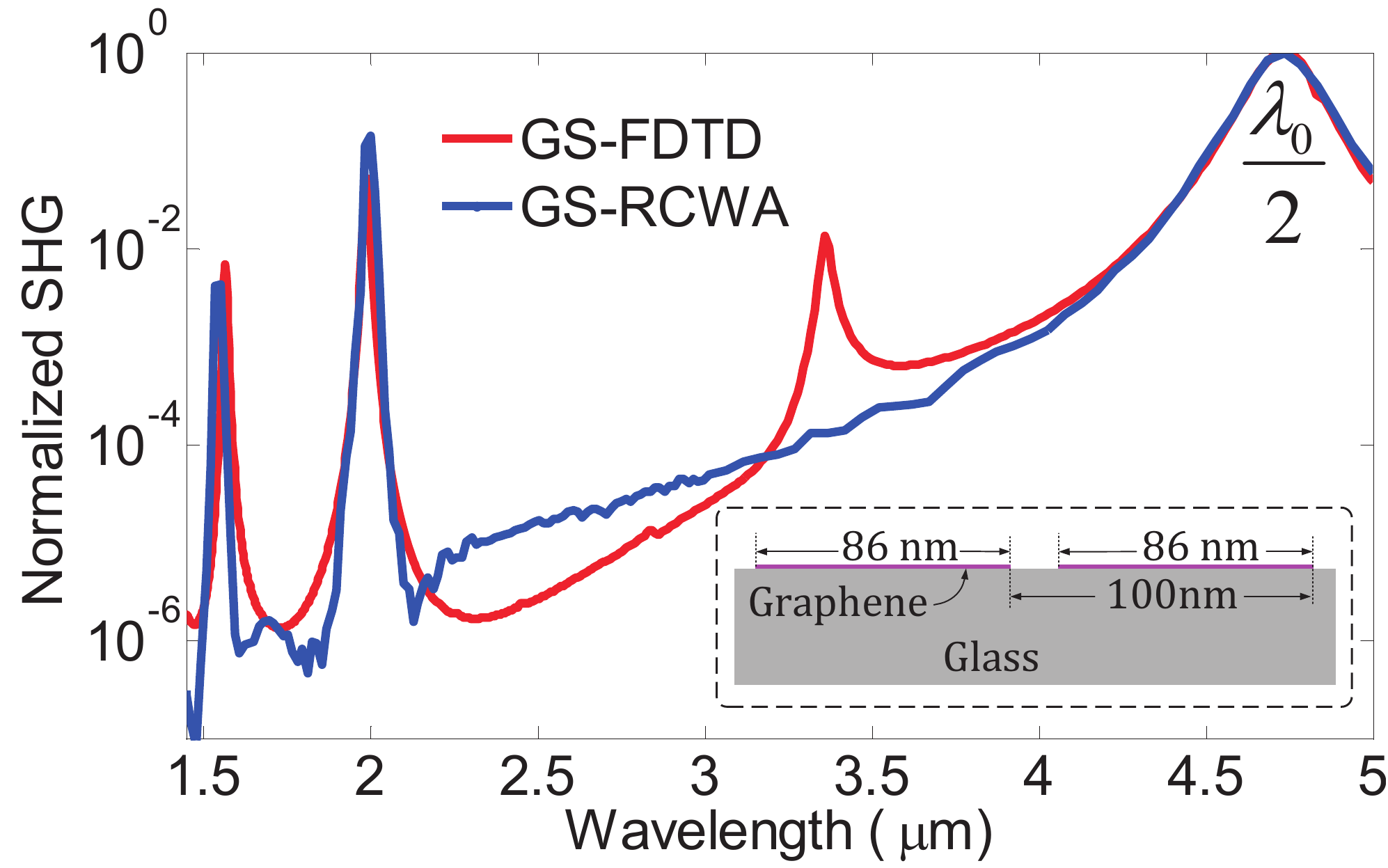}
\caption{Comparison of SHG from graphene grating on a glass substrate, calculated using two
different methods.}\label{fig:nonlinSHG}
\end{figure}

To illustrate the versatility of our method in describing different types of
nonlinearities, we present now the results pertaining to SHG in the graphene grating placed on a
glass substrate and compare them with the predictions of the GS-RCWA method. The conclusions of
this analysis, presented in \figref{fig:nonlinSHG}, show that although in this case the
predictions of the two methods agree to a lesser extent, the amplitude and width of the plasmon
resonances are correctly evaluated by both methods. The differences in the results obtained by
using the two methods are explained by the fact that Fourier-series-expansion methods, such as
RCWA-type methods, show slow convergence when near-fields are calculated. Although these
issues can be circumvented in some cases \cite{wgp15jop}, in the case of SHG in graphene they
still manifest themselves because, unlike the case of THG, which is mainly determined by the
dominant field component, $E_x$, SHG is chiefly determined by the weak, $E_z$ field component. By
contrast, grid-based methods, such as FDTD, are effective at the evaluation of the near-field
distribution, thus, they are usually more accurate in calculating local nonlinear sources.

Our GS-FDTD method has several other appealing features. First, the GS-FDTD method can be used to
model not only periodic structures, but also single scatterers and devices of finite extent.
Equally important, in addition to diffraction problems, GS-FDTD method can also be used to study
much more complicated nonlinear problems, such as light propagation in a nonlinear medium beyond
the paraxial approximation, design of high-Q nonlinear photonic crystal cavities, and radiation
from clusters of nonlinear nanoparticles.

\section{Conclusion}\label{sec:Concl}
In conclusion, we have introduced a novel finite-difference time-domain type method suited to
accurately study optical structures containing dispersive and nonlinear two-dimensional materials.
The dispersive features of these materials are described using a mixture of well-known dispersion
models, such as Debye, Drude, and Lorentz models, whereas their frequency-dependent nonlinear
response is incorporated in our method \textit{via} generalized source currents defined by
the linear near-field. This general setting allows one to study a multitude of nonlinear
processes as one only needs to know the particular dependence of nonlinear currents on the
linear near-field. Importantly, since these nonlinear currents are computed in the
frequency domain, one avoids the calculation of complex time-domain convolution integrals, thus
significantly increasing the computational efficiency of our method. In addition, in order to
illustrate the versatility of the method, we employed it to calculate the second- and
third-harmonic generation in graphene gratings and showed that good agreement with
alternative numerical approaches, such as finite-element method and rigorous coupled-wave
analysis, is achieved.

\section*{Funding} H2020 European Research Council (ERC) (ERC-2014-CoG-648328).

\end{document}